\documentclass[a4paper,12pt]{iopart}

\usepackage{epsfig,amssymb,citesort}

\begin{document}

\title{Electronic structure, magnetism, and disorder in the Heusler compound 
      Co$_2$TiSn }

\author{Hem Chandra Kandpal,$^1$ Vadim Ksenofontov,$^1$\\ 
        Marek Wojcik,$^2$ Ram Seshadri,$^3$
        and Claudia Felser$^1$}

\address{$^1$Institut f\"ur Anorganische Chemie und Analytische Chemie\\
Johannes Gutenberg-Universit\"at, Staudinger Weg 9, 55099 Mainz, Germany\\
kandpal@uni-mainz.de}

\address{$^2$Institute of Physics, Polish Academy of Sciences\\ 
         Al. Lotnikow 32/46, 02-668 Warszawa, Poland}

\address{$^3$Materials Department and Materials Research Laboratory\\
	 University of California, Santa Barbara, CA 93106, USA}

\begin{abstract}
Polycrystalline samples of the Heusler compound 
Co$_2$TiSn have been prepared and studied using bulk techniques (X-ray 
diffraction and magnetization) as well as local probes ($^{119}$Sn M\"ossbauer
spectroscopy and $^{59}$Co nuclear magnetic resonance spectroscopy)
in order to determine how disorder affects half-metallic behavior
and also, to establish the joint use of M\"ossbauer and NMR spectroscopies 
as a quantitative probe of local atom ordering in these compounds. Additionally,
density functional electronic structure calculations on ordered and
partially disordered Co$_2$TiSn compounds have been carried out at a number
of different levels of theory in order to simultaneously understand how the 
particular choice of DFT scheme as well as disorder affect the computed
magnetization. Our studies suggest that a sample which seems well-ordered by
X-ray diffraction and magnetization measurements can possess up to 10\% of 
antisite (Co/Ti) disordering. Computations similarly suggest that even 12.5\% 
antisite Co/Ti disorder does not destroy the half-metallic character of this 
material. However, the use of an appropriate level of non-local DFT is crucial.

\end{abstract}


\maketitle

\section{Introduction}

In recent years, the challenge of creating spintronic devices\cite{WAB01} has
increasingly required, for spin valves and for spin injection,
ferromagnetic materials with high Curie temperatures, high magnetic moments, 
and high spin polarization. These are invariably attributes of half-metallic
ferromagnets.\cite{deGroot_PhysRevLett.50.2024} A half-metal is a ferromagnet 
with a gap in one of the spin directions at the Fermi energy $\epsilon_F$.
Amongst the numerous compounds studied which have this property, the 
Heusler compounds are perhaps the most promising. Many recent investigations on 
bulk \cite{BFJ03,UKK04,UKF05,WFK05,WFK06,BFK06} and 
thin film \cite{GBN05,IOM06,YMI06,BGN06,SBK06} 
Heusler compounds have been carried 
out, and their use in devices has been investigated as 
well.\cite{KTH04,OMS05,SHO06,SMO06,YMI06,OSN06,TIM06} 

A number of electronic structural and magnetic studies have been carried out
on one specific Heusler compound Co$_2$TiSn. For example, Majumdar 
\textit{et al.}\cite{MCS05} have observed a semiconductor-metal transition 
at the Curie temperature of this compound at 350\,K, for which they 
invoked low carrier concentration at the $\epsilon_F$.  
Pierre \textit{et al.}\cite{PSS93} have systematically studied the magnetic 
behavior of Co$_2$TiSn. A number of theoretical studies 
on this compound have also been carried out.\cite{MBS95,LLB05,HHH06,MSN06}
Despite this considerable body of theoretical and experimental work, some 
of the behavior of Co$_2$TiSn remains ambiguous.

The goal of this contribution is two-fold. We use a combination of X-ray 
diffraction and magnetization measurements on a well-annealed (800\,K, 14\,days)
polycrystalline
sample of Co$_2$TiSn to establish that it seems, by these techniques, to
be well-ordered, and a full (integral) moment ferromagnet, which we treat
as a criterion for it being a half-metal.
We then use the \textit{local\/} probes of $^{119}$Sn 
M\"ossbauer spectroscopy and $^{59}$Co spin-echo nuclear magnetic resonance 
spectroscopy to accurately establish the degree of antisite disorder 
in this seemingly well-ordered compound. Finally, we establish that different
levels of density functional theory provide distinctly different results 
regarding whether the compound is half-metallic. Using the highest level
of these computations, we demonstrate that as much as 12.5\% antisite Co/Ti 
disorder does not destroy the half-metallic character.

\section{Experimental and computational methods}

Co$_2$TiSn was prepared by arc-melting the elements under an argon 
atmosphere after many pump/purge steps using 
a 10$^{-4}$\,mbar vacuum. The arc-melting procedure was repeated three times 
to ensure homogeneity. The product was subsequently sealed in an evacuated 
silica tube and annealed at 800\,K for 14\,days. The room temperature X-ray 
diffraction pattern of Co$_2$TiSn was measured on a Bruker D8 instrument 
operated in reflection geometry with a MoK$\alpha_{1,2}$ X-ray source. 
To improve statistics, three datasets were collected and used in the
Rietveld refinement.
SQUID magnetization measurements on the annealed sample were performed on a 
Quantum Design MPMS 5XL magnetometer. The measured saturation magnetic moment 
was 2\,$\mu_B$ per formula unit at 5\,K.

M\"ossbauer measurements on powder samples were performed in the transmission 
geometry using a constant-acceleration spectrometer and a He bath cryostat. 
$^{119}$Sn M\"ossbauer spectrum were measured using a 10\,mCi $^{119}$Sn 
(CaSnO$_3$) source. The \textsc{recoil\,1.03} M\"ossbauer analysis software 
was used to fit the experimental spectrum.\cite{SDB00}  
$^{59}$Co NMR experiments on samples of powdered Co$_2$TiSn were carried out 
at 4.2\,K using a broadband phase-sensitive spin-echo 
spectrometer.\cite{NWJ95} The NMR spectrum was recorded by measuring spin-echo 
intensities. In the final NMR spectrum, the intensities were corrected for the 
enhancement factor and for the usual $\omega^2$ dependence of spectrum 
intensity, to obtain relative intensities that are proportional to the number 
of nuclei with a given NMR resonance frequency. The external magnetic field 
was zero, and a constant excitation RF field was used.\cite{BBV01}

We have used a combination of four different first principles density
functional theory codes: The full-potential linear augmented plane wave
code \textsc{Wien2k}\cite{BSM01} and the full-potential linear muffin 
tin orbital (\textsc{fplmto}) method,\cite{SSa92} and for comparison, 
the LMTO method within the atomic sphere approximation 
(\textsc{lmto-asa})\cite{JAn00} and the spin-polarized relativistic 
Korringa-Kohn-Rostroker \textsc{sprkkr}\cite{Ebe99} method as well. These 
latter methods approximate the potential within a crystal to be a summation of 
spherical potentials over the atomic sites (\textsc{sprkkr}) and atomic and 
interstitial sites (\textsc{lmto-asa}). The experimental cell parameter was 
used in all the calculations.

The exchange-correlation energy functional was evaluated within the local
density approximation (LDA), using the von-Barth-Hedin \cite{vBH72} as well
as generalized gradient approximation (GGA), using the
Perdew-Burke-Ernzerhof \cite{PBE96} parametrization. Muffin-tin radii
(RMTs) were taken in the range 2.3 to 2.36\,$a_{Bohr}$ 
 for all the atoms, and this resulted in nearly 
touching spheres. Self-consistent calculations employed a grid of 455 
irreducible $k$ points on a $25\times25\times25$ mesh in the irreducible wedge
of the Brillouin zone. This number of irreducible $k$ points was found to be
sufficient for convergence. The energy convergence criterion was set to 
$10^{-5}$. Charge convergence was monitored concurrently.

\section{Results and discussion}

\subsection{Experiments}

%%%%%%%%%%%%%%%%%%%%%%%%%%%%%%%%%%%%%%%%%%%%%%%%%%%%%%%%%%%%%%%%%
\begin{figure}[h] 
\centering\epsfig{file=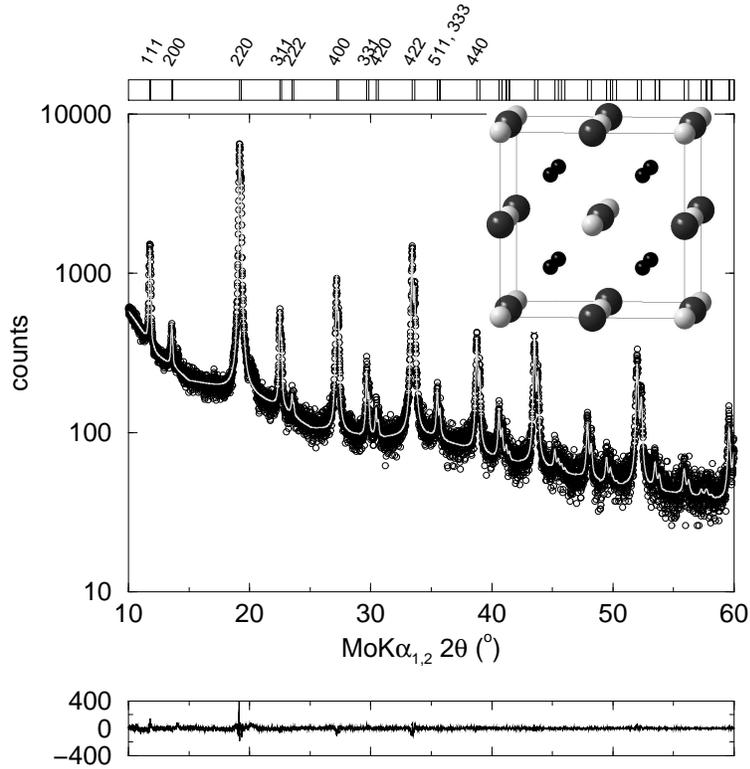, width=10cm}
\caption{Mo\textit{$K_\alpha$} X-ray powder diffraction pattern of an 
annealed Co$_2$TiSn sample, plotted on a semilog scale. Points are data, and 
the gray line is the Rietveld fit. The difference profile is also displayed
in the panel below, in linear counts. The inset is the Heusler
crystal structure showing 
Co atoms (small black spheres) at 8$a$ $(\frac 1 4 \frac 1 4 \frac 1 4)$,
Ti atoms  (small light-gray spheres) at 4$a$ $(0 0 0)$, and
Sn atoms  (large dark-gray spheres) at 4$b$ $(\frac 1 2 \frac 1 2 \frac 1 2)$.
Vertical lines at the top of the plot are the expected $\alpha_1$ and 
$\alpha_2$ peak positions. The low angle peaks are indexed.
}
\label{XRAY} 
\end{figure}
%%%%%%%%%%%%%%%%%%%%%%%%%%%%%%%%%%%%%%%%%%%%%%%%%%%%%%%%%%%%%%%%% 

Figure\,\ref{XRAY} displays the Mo-K$\alpha$ X-ray powder diffraction pattern
of the annealed Co$_2$TiSn sample. Experimental data are displayed as points.
The data were subject to refinement using the Rietveld method as implemented
in the \textsc{xnd} code.\cite{xnd}
The gray line is the Rietveld fit ($R_{Bragg}$ = 3.7\%) to the cubic 
$Fm\overline 3 m$ Heusler structure with a cell parameter that refined to  
6.0718(3)\,\AA. Since the atomic number of Sn ($Z$ = 50) is well distinguished
from the atomic numbers of Ti ($Z$ = 22) and Co  ($Z$ = 27), we performed
a number of simulations where the Sn site was partially occupied by these 
lighter atoms. These simulations suggested that Sn is fully ordered in this
compound. The small $Z$ difference between Co and Ti did not allow for their
relative occupancies in the two sites to be refined, so that in the refinement 
model, their occupancies were fixed to one corresponding to 8.8\% antisite 
Co/Ti disorder as suggested by the other \textit{local\/} probes presented 
here. The refined isotropic thermal parameters for all atoms were somewhat 
large but reasonable, in the range of $B$ = 1.4 to 1.7\,\AA$^2$. 

%%%%%%%%%%%%%%%%%%%%%%%%%%%%%%%%%%%%%%%%%%%%%%%%%%%%%%%%%%%%%%%%%
\begin{figure}[h]
\centering \epsfig{file=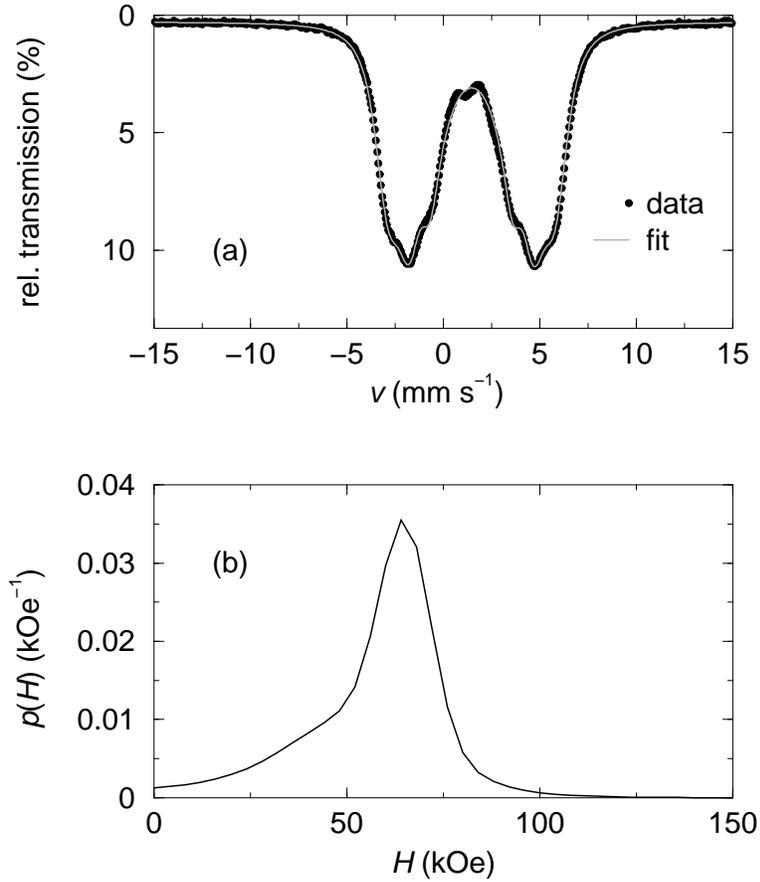, width=10cm}
\caption{(a) $^{119}$Sn M\"ossbauer spectrum of annealed Co$_2$TiSn sample
recorded at 295\,K and (b) the hyperfine magnetic field distribution on Sn 
atoms in Co2TiSn. The spectrum is fit using a single isomer shift but
assuming the following distribution of hyperfine fields and their relative
intensities: 65.9(1)\,kOe (49\%),  56.4(8)\,kOe (39\%), and 20(1)\,kOe   
(12\%).} 
\label{MOSS} 
\end{figure}
%%%%%%%%%%%%%%%%%%%%%%%%%%%%%%%%%%%%%%%%%%%%%%%%%%%%%%%%%%%%%%%%%

The $^{119}$Sn M\"ossbauer spectrum of annealed sample of Co$_2$TiSn measured
at room temperature is shown in figure~\ref{MOSS}(a). To fit the spectrum  
a magnetic hyperfine field distribution model was employed. The Co$_2$TiSn 
spectrum can be decomposed into three sub-spectra with the same isomer shift 
IS = 1.48(2)\, mms$^{-1}$ and zero quadrupole splitting. The partial 
intensities and hyperfine magnetic fields of the three sub-spectra are 
provided in the caption of figure\,\ref{MOSS}. A resolved hyperfine
structure is revealed in the distribution $p(H)$ displayed 
figure~\ref{MOSS}(b). The asymmetrical distribution has a maximum at
65.9(1)\,kOe and small intensity at zero value of the hyperfine field. 
Note that the hyperfine field density distribution curve for a completely
ordered compound should contain only one symmetrical peak. The asymmetry 
in the $p(H)$ distribution as a function of the hyperfine field suggests 
partial disordering of the environment around Sn. To complete the 
interpretation, we turn to $^{59}$Co NMR spectroscopy.

%%%%%%%%%%%%%%%%%%%%%%%%%%%%%%%%%%%%%%%%%%%%%%%%%%%%%%%%%%%%%%%%%
\begin{figure} 
\centering \epsfig{file=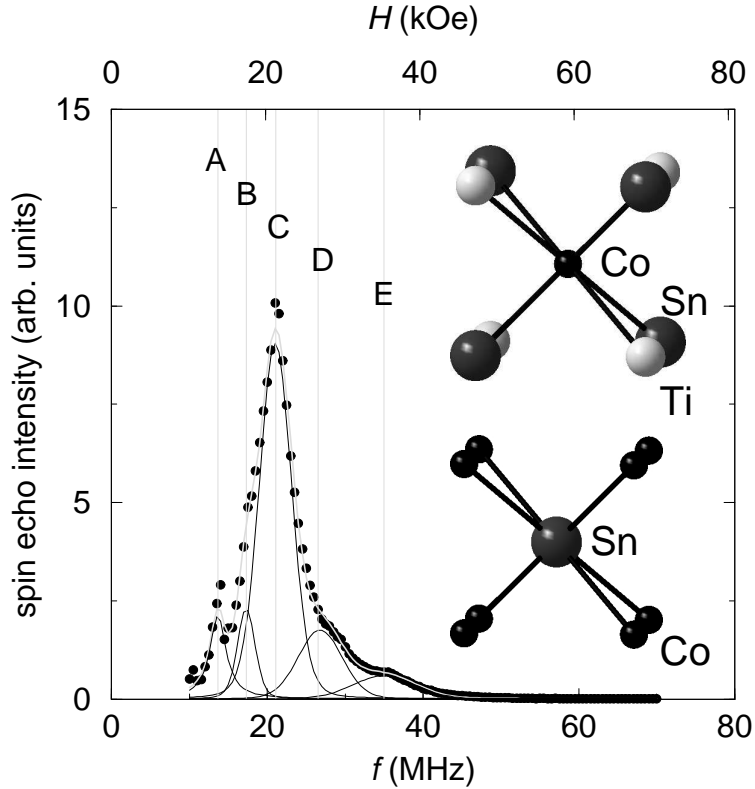, width=10cm}
\caption{$^{59}$Co NMR spectrum of Co$_2$TiSn. The data could be fit using 
five Gaussian peaks whose centers are indicated by vertical gray lines. The 
resonance frequencies (relative intensities) of the five peaks (labeled A 
through E) are: $f_A$ = 13.7(3)\,MHz (10\%), $f_B$ = 17.3(4)\,MHz (8\%), 
$f_C$ = 21.1(3)\,MHz (60\%), $f_D$ = 27(3)\,MHz (15\%), and 
$f_E$ = 35(4)\,MHz (7\%). The upper abscissa displays the equivalent hyperfine 
fields. The insets show the (4Ti + 4Sn) coordination of Co and the 8Co 
coordination of Sn.}
\label{NMR} 
\end{figure}
%%%%%%%%%%%%%%%%%%%%%%%%%%%%%%%%%%%%%%%%%%%%%%%%%%%%%%%%%%%%%%%%%

The $^{59}$Co NMR spectrum of Co$_2$TiSn acquired at 5K is presented in 
figure\,3. The resonance frequencies $f$ are related to the hyperfine fields 
(indicated on the upper abscissa) through the gyromagnetic ratio, 
$g$ = 1.0103\,kHz\,Oe$^{-1}$. The spectrum can be decomposed into five 
Gaussian peaks A through E with the parameters described in the caption. 
The dominant line in the spectrum at frequency $f_C$ = 21.1(3)\,MHz is unsplit 
in agreement with the cubic structure of Co$_2$TiSn. Two satellite lines 
located at $f_D$ = 27(3)\,MHz and $f_E$ = 35(4)\,MHz correspond to Co atoms 
experiencing higher hyperfine fields in comparison to the main line. 
We consider the main line as originating from Co atoms in
ordered stoichiometric surroundings, whereas the satellites stem from the
Co positions with Co atoms in their first coordination sphere. The first 
coordination sphere of Co atoms in the fully ordered structure would be
4 Ti and 4 Sn atoms. Statistical mixing of Ti and Co atoms should obey the 
expression for the probability to find $n$ impurity atoms from amongst 
$N$ neighbors using a binomial distribution: 

\[W_n = \frac{N!}{n!(N-n)!}(1-x)^nx^{N-n} \]

\noindent where $x$ is
fraction of ``extrinsic'' atoms. For example, the probability corresponding to 
a single extra Co atom substituting a Ti would correspond to $N$ = 4 (for
the four usual Ti neighbors) and $n$ = 1 (for the Co atom substituent).  
The distribution describing the
probability of observing the ``undisturbed'' first coordination sphere suggests
7.8\% of Co atoms substituting Ti atoms. The resonance line at $f_D$ = 
 27(3)\,MHz originates from a Co atom with one Co atom substituting one of 
the 4 Ti in its first coordination shell, and at $f_E$ =  35(4)\,MHz, 
the Co atoms being probed has two Co atoms substituting for Ti.

The second coordination sphere of Co atoms comprises six Co atoms. 
Substitution of the Co atoms in the second coordination sphere by nonmagnetic 
Ti atoms should decrease the hyperfine magnetic field and hence the resonance 
frequency. The resonance line at $f_B$ = 17.3(4)\,MHz is attributed to a 
Co having one of its six Co in the second coordination sphere being replaced 
by Ti, and the line at $f_A$ = 13.7(3)\,MHz is then attributed to a Co with 
two Ti atoms, substituting for two Co atoms in its second coordination sphere.
The binomial distribution model gives the amount of Ti atoms
substituting Co atoms to be 8.8\%. Taking into account the amount of Co
atoms substituting Ti atoms, the final composition describing the antisite
disordering, can be approximately written:
[Co$_{(2-0.09)}$Ti$_{(0.09)}$][Ti$_{(1-0.09)}$Co$_{(0.09)}$]Sn or more
concisely as (Co$_{1.91}$Ti$_{0.09}$)(Ti$_{0.91}$Co$_{0.09}$)Sn.

With an understanding of disorder from NMR, we can return to the X-ray 
diffraction in order to understand why it is unable to discern the disorder.
The exchange of Ti with Co atoms on both $8a$ positions, as suggested by NMR,
is indicative of the structure being partially DO$_3$-like. Because both 
the usual Heusler (L2$_1$) and the DO$_3$ structure type have the same space
group ($Fm\bar 3m$), and because the atomic numbers of Ti and Co 
are not well separated, 
the determined composition (Co$_{1.91}$Ti$_{0.09}$)(Ti$_{0.91}$Co$_{0.09}$)Sn 
is not easily distinguished by X-ray diffraction from pristine Co$_2$TiSn.

The site assignment of Sn atoms follows from the statistical analysis of
intensities obtained from M\"ossbauer spectroscopic measurements.
First coordination sphere of Sn atoms comprises eight Co atoms. Partial 
substitution of Co atoms by Ti atoms should diminish the
hyperfine magnetic field on Sn atoms. This effect clearly follows from the
hyperfine field distribution presented in the caption of figure\,\ref{MOSS}.
The sub-spectrum with a hyperfine field of 65.9(1)\,kOe can be assigned to 
the ``undisturbed'' configuration of Sn atoms. The sub-spectrum with the
reduced hyperfine field of 56.4(8)\,kOe corresponds to Sn with seven
Co and one  Ti neighbors. The part of distribution with a hyperfine field of 
20(1)\,kOe indicates the further increase in the amount of Ti atoms 
substituting Co atoms in the first coordination sphere of Sn; six Co and
2 Ti. The binomial distribution then suggests 8.6\% of Ti atoms substituting Co
on average which is in excellent agreement with the 8.8\% proposed based on
the $^{59}$Co NMR experiment. The composition   
(Co$_{1.91}$Ti$_{0.09}$)(Ti$_{0.91}$Co$_{0.09}$)Sn is therefore consistent
with the $^{119}$Sn M\"ossbauer data as well.

\subsection{Computation}

%%%%%%%%%%%%%%%%%%%%%%%%%%%%%%%%%%%%%%%%%%%%%%%%%%%%%%%%%%%%%%%%%
\begin{figure} 
\centering \epsfig{file=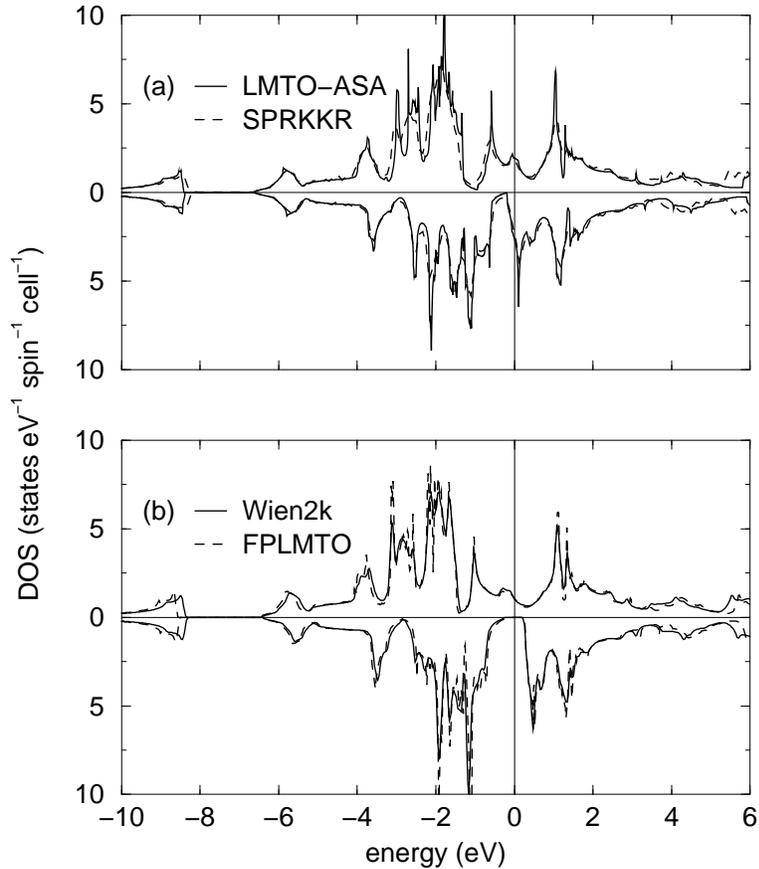, width=10cm}
\caption{(a) Densities of state of ordered Co$_2$TiSn obtained using the 
\textsc{lmto-asa} and \textsc{sprkkr} codes. (b) Densities of state of 
ordered Co$_2$TiSn obtained using the \textsc{Wien2k} and \textsc{fplmto} 
codes. 0 on the energy axis is $\epsilon_F$. 
All the calculations used the generalized gradient approximation.} 
\label{DOS1} 
\end{figure}
%%%%%%%%%%%%%%%%%%%%%%%%%%%%%%%%%%%%%%%%%%%%%%%%%%%%%%%%%%%%%%%%%

Densities of state of ordered Co$_2$TiSn obtained using the different
methods are shown in figure\,\ref{DOS1}. It is seen that the methods
using spherical potentials (\textsc{lmto-asa} and \textsc{sprkkr}) fail 
to obtain the correct, measured full moment, which we take here 
to mean a half-metallic ground state. Panel\,(a) 
of this figure shows that within these computational schemes, minority spin 
state are occupied. The full potential 
schemes embodied in the \textsc{Wien2k} and \textsc{fplmto} codes however do 
correctly obtain a 
minority gap in this compound, and the full, measured magnetic moment of
2\,$\mu_B$ \textit{per\/} formula unit. Very little, if any, difference is
seen between the two spherical potential codes, and between the two
full-potential methods. The calculated moments using different codes, and
using LDA and GGA are summarized in table~\ref{DFT}. The calculated total 
magnetic moments are in the range from 0.84 to 2.00\,$\mu_B$. It is also
noted that in addition to using full-potential methods, gradient corrections
(GGA) help to obtain the correct full moment electronic 
structure description of this compound.

%%%%%%%%%%%%%%%%%%%%%%%%%%%%%%%%%%%%%%%%%%%%%%%%%%%%%%%%%%%%%%%%%
\begin{table}
\caption{Magnetic moments of ordered Co$_2$TiSn calculated using different
schemes.}
\centering 
\begin{tabular}{l|ll} 
\hline 
Code & LDA moment ($\mu_B$) & GGA moment ($\mu_B$)  \\ 
\hline 
\textsc{lmto-asa} & 0.84  & 1.40\\ 
\textsc{sprkkr}   & 1.11  & 1.55\\ 
\textsc{Wien2k}   & 1.99  & 2.00\\
\textsc{fplmto}   & 1.99  & 2.00\\ 
\hline \end{tabular} 
\label{DFT} 
\end{table}
%%%%%%%%%%%%%%%%%%%%%%%%%%%%%%%%%%%%%%%%%%%%%%%%%%%%%%%%%%%%%%%%%

%%%%%%%%%%%%%%%%%%%%%%%%%%%%%%%%%%%%%%%%%%%%%%%%%%%%%%%%%%%%%%%%%
\begin{figure} 
\centering \epsfig{file=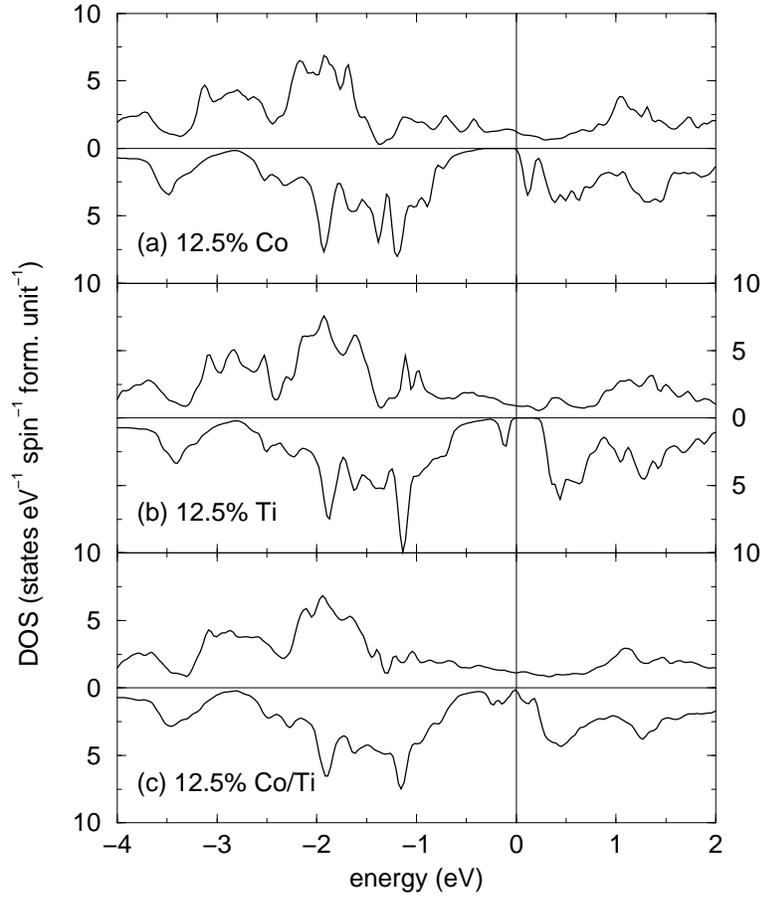, width=10cm}
\caption{Densities of state of disordered Co$_2$TiSn obtained using
the \textsc{Wien2k} code: (a) 12.5\% Co substituting Ti; 
                          (b) 12.5\% Ti substituting Co;
                          (c) 12.5\% Co/Ti swap.} 
\label{DOS2} 
\end{figure}
%%%%%%%%%%%%%%%%%%%%%%%%%%%%%%%%%%%%%%%%%%%%%%%%%%%%%%%%%%%%%%%%%

In agreement with our NMR and M\"ossbauer experiments, the most probable
defects in Co$_2$TiSn are Co-Ti swap, which give rise to the 
general formula (Co$_{2-z}$Ti$_z$)(Ti$_{1-z}$Co$_z$)Sn, with $z$ 
in this case being close to 0.09. Two other kinds of disorder \cite{PCF04} 
can be considered: A Co-antisite where a Ti atom is replaced by  Co, 
and a Ti-antisite where a Co atom is replaced by Ti. 
We do not consider any disordering of Sn since neither the local probes, nor
X-ray diffraction give any suggestion of it. We have considered all three cases 
of disordering in band structure calculations, with a disordering rate of 
12.5\%. This value was chosen because it is easily implemented in supercells 
involving doubling lattice parameters in all three directions. It is also
close to what is experimentally observed. It should be noted that all three
modes of disorder require require lowering of symmetry from cubic.

Figure\,\ref{DOS2} shows the \textsc{Wien2k}-GGA densities of state, scaled 
to one Co$_2$TiSn formula unit, for the three cases of disordering. In order
to focus on the gap in the minority spin direction, the data are displayed 
in a small window of energy around $\epsilon_F$.
It is seen that in all cases, the minority gap at $\epsilon_F$ 
(half-metallic character) is retained. All three modes of disorder results 
in new states being created in the minority gap of pure Co$_2$TiSn. The
majority states are nearly unaffected by the disorder. Excess Co substituting 
for Ti is seen to create states above $\epsilon_F$, whereas excess Ti 
substituting for Co is seen to create states below $\epsilon_F$. The swapping
of Co and Ti does a little of both, and the gap as a consequence, is 
almost lost. In general, these Heusler compounds seem to be robust half-metals.
The calculations refer to zero Kelvin, and it can be expected that smearing 
of states at finite temperature will further diminish the gap, and at least for
the last case (c) of the Co/To swap; disorder in conjunction
with finite temperatures would be deleterious for half-metallic 
behavior.

\section{Conclusions}

This work allows a number of conclusions to be drawn from the combination
of experiments and computation on an important Heusler compound. From 
experiments, we observe that compounds that seem to be ordered, from X-ray 
diffraction, and from magnetization measurements, can through \textit{local\/}
probes such as M\"ossbauer and NMR be found to possess significant and 
quantifiable antisite disorder. In this particular case, the precise nature
of the disorder is consistent with approximately 9\% of Co and Ti exchanging
their lattice sites. The power of M\"ossbauer and NMR used together in 
establishing local disorder has been demonstrated.

Computationally, the very interesting result is demonstrated that different
implementations of density functional theory provide distinctly different
results. To correctly reproduce the half-metallic ground state of Co$_2$TiSn,
both non-local descriptions of the exchange correlation functional (GGA) as
well as, more importantly, non-spherical potentials are required to be used
in the calculations. Thus, it is only the full-potential methods that are
able to correctly represent the electronic structure of Co$_2$TiSn. 
In agreement with experiment, the system can accommodate 
quite a large degree of antisite disorder without loosing its half-metallic 
character.

\ack {This work is financially supported by the DFG (project TP1 and TP7 in research
group FG 559). RS gratefully acknowledges the National Science
Foundation for support through a Career Award (DMR04-49354).}

\medskip

\providecommand{\newblock}{}

\end{document}